\documentstyle[graphicx,aps]{revtex}
\tightenlines

\begin{document}


\title{The dynamics of a strongly driven two component Bose-Einstein
Condensate.}
\author{G.~L.~Salmond$^1$,C.~A.~Holmes$^2$ and G.~J.~Milburn$^3$}
\address{$^1$Center for Laser Science, Department of Physics,\\
$^2$Center for Mathematical Physics,\\
Department of Mathematics,\\
$^3$Center for Quantum Computer Technology,\\
Department of Physics,\\
 The University of Queensland,
Queensland 4072 Australia.}
\date{\today}
\maketitle

\begin{abstract}
We consider a two component Bose-Einstein condensate
in two spatially localized modes of a double well potential, with
periodic modulation of the tunnel coupling between the two modes. We
treat the driven quantum field using a two mode expansion and define
the quantum dynamics in terms of the Floquet Operator for the time periodic
Hamiltonian of the system.  It has been shown that the corresponding
semiclassical
mean-field dynamics can exhibit regions of regular and chaotic motion.
We show here that the quantum dynamics can exhibit dynamical tunneling
between regions of regular motion, centered on fixed points
(resonances) of the semiclassical dynamics.
\end{abstract}

\section{Introduction}

Since the first experimental realization of a Bose condensed gas of
atoms in 1995~\cite{andersonetal} there has been considerable interest
in the properties of Bose-Einstein Condensates (BEC's). Here we are
interested in the properties of a BEC confined to the two spatially
localized regions of a double well potential, and subjected to periodic
modulation in such a way that the tunnel coupling between the two wells
is periodically modulated.  In this way it has been shown that the
semiclassical mean-field dynamics can exhibit regions of chaotic
motion~\cite{spie,abdullaev}.

The primary goal of this paper is to develop an understanding of the
quantum dynamics, beyond mean-field theory, of a dynamical system in a
classically chaotic regime. In particular we seek quantum features of
the dynamics of strongly driven BEC's that are not obtained in
mean-field theory. Over the last two decades there have been numerous
studies of nonlinear systems with a finite number of degrees of
freedom; a topic that is often referred to as `quantum chaos', see for
example~\cite{stockmann,haake}. In this paper however we attempt to
address the quantum dynamics of a driven quantum field. In the
mean-field limit, an effective classical field (a system of an infinite
number of degrees of freedom) is used  to describe the condensate. The
response of the system to external periodic forcing would then be described by
the driven Gross-Pitaevskii equation.
In the hydrodynamic limit the mean field description
would look very much like a forced nonlinear fluid dynamics model~\cite{fluid}.
A similar system of equations also arise in periodically modulated nonlinear
optical fibers~\cite{optical}. Even without external
time dependent forcing nonlinear field equations of the kind considered
here exhibit
dynamical instabilities leading to a host
of interesting phenomenon for both the integrable
case (e.g. solitons) and the non-integrable case (e.g.
turbulence)~\cite{Cai,mclaughlin}.
The response of condensates to forcing can encompass a large variety of
physical situations including vortex formation in response to
stirring~\cite{Brand2001},
and rotating condensates in an anisotropic trap~\cite{Sinha2001}.
Adhikari~\cite{Adhikari2001}
recently conducted a numerical study, in mean field theory,
of two couple condensates with modulation of the trap frequency
and also modulation of the nonlinear term due to hard sphere collisions.
Gardiner et.al.~\cite{gardiner} considered a BEC in a modulated
periodic potential.

We will consider two condensates with a tunnel coupling
periodically modulated in time. In a fully quantum treatment
the dynamics is given by a nonlinear equation
for the field operator. We do not know in general how to deal with such
systems. In this paper we adopt the usual method of a finite mode
expansion, which effectively reduces both the classical and quantum
systems to systems with few degrees of freedom. Even at this crude
level of approximation however we can find fundamental differences
between the mean-field predictions and those of the quantum
description.

In this paper we focus on two tunnel-coupled BEC's in a double well
potential, including the nonlinear self energy term, with periodic
modulation of the tunnel coupling. A similar model has recently been
studied by Abdullaev and Kraenkel~\cite{abdullaev} in the mean-field
limit. They also included modulation of the energy difference between
the two condensates as well as dissipation.  They showed that, in
mean-field theory,  the system could exhibit chaotic oscillations of
the relative population difference between the two wells.  Recently,
Elyutin and Rogovenko have treated a periodically driven double well
model via the Gross-Pitaevskii equation~\cite{elyutin}.  There are many
studies of a BEC in a time independent double well
potential~\cite{Javanainen1986,Dalfovo1996,Andrews1997,Smerzi1997,milburn1997,Raghavan1999,elenaetal}.
We will use the model of reference~\cite{milburn1997} which is valid
for small condensates (few atoms, i.e. $N<1000$), where one expects
quantum departures from mean-field theory to be more significant.  We
make a two mode approximation for the double well
system~\cite{milburn1997,spekkens99,cirac98}, with weak interwell
tunneling.  As particle number is a constant of the motion at zero
temperature, we can use an approach based on the two mode bosonic
realization of the SU(2) algebra.  We use the Schr\"{o}dinger picture
and an angular momentum model for the quantum system in SU(2) dynamics.
We find a parameter regime in which an initial atomic coherent state
tunnels from one fixed point (resonance) to another.  This behavior is
analyzed from a Floquet state perspective and reveals an interesting
feature of parity of the Floquet operator eigenstates.  

The many body Hamiltonian is specified in section~\ref{ham}, and our
two mode approximation to it is given.  Section~\ref{eqofm} outlines
the models and the corresponding equations of motion for the two mode
system.  The semiclassical and quantum results are discussed in
sections~\ref{semiclassicaldynamics} and~\ref{qm} respectively,
followed by our concluding remarks in section~\ref{conclusion}.

\section{The Hamiltonian}
\label{ham}

The many-body Hamiltonian for an atomic BEC confined in a potential
$V({\mathbf{r}},t)$ is given by~\cite{bec-griffin},
\begin{equation}
\hat{H}(t)=\int d^3{\mathbf{r}} \left[ \frac{\hbar^2}{2m}
\nabla\hat{\psi}^\dagger\cdot\nabla\hat{\psi}
+\hat{\psi}^\dagger V({\mathbf{r}},t) \hat{\psi}
+ \frac{U_0}{2} \hat{\psi}^\dagger\hat{\psi}^\dagger
\hat{\psi}\hat{\psi} \right]
\label{manybodyham}
\end{equation}
where $m$ is the boson mass, $U_0=4\pi\hbar^2a/m$ measures the
strength of the two body interaction and $a$ is the s-wave
scattering length.
$\hat{\psi}^\dagger=\hat{\psi}^\dagger({\mathbf{r}},t)$ and
$\hat{\psi}=\hat{\psi}({\mathbf{r}},t)$ are the Heisenberg
creation and annihilation operators of particles at position
${\mathbf{r}}$. $V({\mathbf{r}},t)=V_x({\mathbf{r}})(1+\epsilon \cos
\omega_Dt)+V_y({\mathbf{r}})+V_z({\mathbf{r}})$, and $\epsilon < 1$.
$V({\mathbf{r}},t)$ is a time
dependent potential, parameterised by the modulation strength
$\epsilon$, and the driving frequency $\omega_D$.  Throughout this
paper $\epsilon=0$ implies non-driven
dynamics and $\epsilon>0$ corresponds to driven (modulated) dynamics.

We will follow the model of reference \cite{milburn1997} in
which the trap potential is taken to be a symmetric double well
in the $x$-direction and harmonic in the $y,z$-directions,
\begin{equation}
V({\mathbf{r}})=b(x^2-q_0^2)^2+\frac{1}{2}m \omega_t^2(y^2+z^2)
\end{equation}
where $\omega_t$ is the trap frequency in the $y-z$ plane,
 and where $b$ gives the strength of the confinement in the
$x$-direction. The potential has stable fixed points at
${\mathbf{r}}_1=+q_0{\mathbf{x}},\ \ {\mathbf{r}}_2=-q_0{\mathbf{x}}$
near which the linearised motion is harmonic with frequency
$\omega_0=q_0(8b/m)^{1/2}$. For simplicity we will set
$\omega_t=\omega_0$.  The length scale is set by the rms
position fluctuations, $r_0$, in the ground state
of the harmonic potential near the fixed points;
$r_0=\sqrt{\hbar/2m\omega_0}$.
In terms of this parameter the barrier height separating the two wells may be
written $B=(\hbar\omega_0/8)(q_0/r_0)^2$.

We assume the potential is such that there are two nearly degenerate single
particle
energy eigenstates below the barrier and expand the field in
terms of two localised single particle states,
$u_d({\mathbf{r}})$, at each stable minima;
\begin{equation}
u_{d}({\mathbf{r}})=u_0({\mathbf{r}}-{\mathbf r}_d), \; \; d=1,2
\label{wellmode}
\end{equation}
where $u_0({\mathbf{r}})$ is defined to be the normalized
single-particle ground state mode of the harmonic potential
near each stable fixed point, ${\mathbf r}_d$, each with energy $E_0$.
The details can be found in \cite{milburn1997}. The energy eigenstates
of the double well may then be approximated as
symmetric ($+$) and antisymmetric ($-$) combinations of the localised states
with energy eigenvalues given in first order perturbation theory by
$E_\pm= E_0\pm\hbar\Omega/2$ where the energy level splitting determines
the tunnelling frequency and is given by
\begin{equation}
\Omega=\frac{3}{8}\omega_0\frac{q_0^2}{r_0^2}e^{-q_0^2/(2 r_0^2)}
\end{equation}
As discussed in \cite{milburn1997} the validity of the two mode
approximation requires $\Omega/\omega_0\ <<\ 1$.

The effect of the modulation is to directly
modulate the parameter $b$ so that it becomes $b(1+\epsilon\cos\omega_Dt)$.
This has two important effects; it modulates the harmonic
frequency around each fixed point and it modulates the tunnelling
frequency, so that
these quantities become time dependent $\omega_0(t),\  \Omega(t)$, where
\begin{eqnarray}
\omega_0(t) & = & \omega_0\sqrt{1+\epsilon\cos\omega_D t}\\
\Omega(t) & = & \Omega_0(1+\epsilon\cos\omega_D t)\exp\left
[-\frac{q_0^2}{2 r_0^2}(\sqrt{1+\epsilon\cos\omega_D t}-1)\right ]
\label{fullomegaoft}
\end{eqnarray}
The two mode approximation then results in the following
Hamiltonian~\cite{milburn1997},
\begin{eqnarray}
\hat H_2(t)&=&\hbar \omega_0(t)(c_1^\dagger c_1+c_2^\dagger c_2)
+\frac{\hbar\Omega(t)}{2}(c_1c_2^\dagger+c_1^\dagger c_2)
\nonumber \\
&+&\hbar{\kappa}\left
((c_1^\dagger)^2c_1^2+(c_2^\dagger)^2c_2^2\right ),
\label{semiclassicalham}
\end{eqnarray}
where
\begin{equation}
c_d(t)=\int
d^3{\mathbf{r}}u^*_d({\mathbf{r}})\hat{\psi}({\mathbf{r}},t),
\label{ccdagger}
\end{equation}
and similarly for $c^{\dagger}_d(t)$, $d=1,2$.  $c_1^\dagger, c_{1}$
and $c_2^\dagger, c_{2} $ are creation and annihilation operators for
atoms in the local modes in wells 1 and 2 respectively. The remaining
parameters in the
Hamiltonian are as follows, $\kappa$ represents the particle-particle
interaction and $\Omega(t)$ measures the tunnelling rate of atoms between
wells 1 and 2. We will assume that the system remains in a total number
eigenstate,
with eigenvalue $N$.
We first note that we will be interested in modulation frequencies that are
of the order of
the  unmodulated tunnelling frequency, $\omega_D\approx \Omega_0$. The
condition
for the validity of the two mode approximation then ensures that $\omega_D\
<<\ \omega_0$.
 We can then replace the first term in the Hamiltonian by a slowly varying
number $\hbar\omega_0(t)N$, which has no influence on the quantum dynamics.
The
temporal modulation of the second term, which is responsible for tunnelling,
has a strong influence on the dynamics as we show.

If $\epsilon$ is small enough we can approximate the modulated tunnelling
frequency
by
\begin{equation}
\Omega(t)\approx\Omega_0(1+\epsilon\cos\omega_D
t)\exp[-\gamma\epsilon\cos\omega_D t]
\end{equation}
where $\gamma=q_0^2/(4 r_0^2)$. While $\epsilon$ may be small we cannot
necessarily claim
that $\gamma\epsilon$ is small.  We may expand the approximate expression for
$\Omega(t)$ using Fourier series,
\begin{equation}
\Omega(t)\approx\Omega_0(1+\epsilon cos\omega_Dt)\sum_{n=-\infty}^\infty
I_n(-\gamma\epsilon)e^{-in\omega_D t}
\end{equation}
where $I_n(x)$ is a Bessel function.
In general we see that the direct modulation of the potential leads to a
non-harmonic
modulation of the tunnelling frequency. In what follows we will neglect the
higher harmonics
of the modulation frequency, and consider the time dependence of
$\Omega(t)$ to be simply
\begin{equation}
\bar{\Omega}(t)=\Omega_0(1+\epsilon\cos\omega_Dt)
\end{equation}
While we have assumed that $\epsilon$ is small, this constraint is not as
strong
as one might expect. When we study the ratio of the approximate
modulated tunnelling to the complete expression given in
Eq.(\ref{fullomegaoft}),
$\bar{\Omega}(t)/\Omega(t)$,
we find that for appropriate values of $q_0/r_0$, the ratio is close to
unity even for
values of $\epsilon$ as large as $0.3$.  Henceforward we will assume the
simpler harmonic modulation of $\Omega(t)=\Omega_0(1+\epsilon\cos\omega_D t)$.

The separation of the higher excited states from the nearly degenerate
ground states is
of the order of $\omega_0$. As the validity of the two mode approximation
requires that $\Omega_0\ <<\ \omega_0$,
and we assume that the driving frequency is of the
order of the  tunnelling frequency.  The driving frequency is very far from
resonance with higher excited states
of the double well and we can continue to make a two level approximation
when the modulation
is present. Of course when the tunnelling and the modulation is slow, the
interesting dynamics
will occur on a long time scale and one may legitimately ask if dissipation
should not
be included. In this paper however we will neglect the effect of dissipation
and focus only on the nonlinear quantum dynamics. Our results are thus
strictly only valid for zero
temperature. The validity of an expansion in terms of localised single particle
states places a restriction on the atomic number~\cite{milburn1997}. We
are thus
only concerned with very small condensates at temperatures low enough that
the non-condensate fraction,
and the resulting dissipation, can be neglected. It is only in
this regime that quantum (rather than semiclassical)
nonlinear effects in driven BEC's could be observed.

Using the generators for SU(2) dynamics,
\begin{eqnarray}
\hat{J}_{x}&=&\frac{1}{2}(c_2^\dagger c_{2}-c_1^\dagger c_{1}),
\label{jjj1} \\
\hat{J}_{y}&=&\frac{i}{2}(c_2^\dagger c_{1}-c_1^\dagger c_{2}),
\label{jjj2} \\
\hat{J}_{z}&=&\frac{1}{2}(c_1^\dagger c_{2}+c_2^\dagger c_{1}),
\label{jjj3}
\end{eqnarray}
the two mode Hamiltonian may be written,
\begin{equation}
\hat{H}_2(t)=\hbar\Omega(t)\hat{J_{z}}+2\hbar\kappa\hat{J_{x}^{2}}.
\label{twomodehamshort}
\end{equation}
It is worth noting the symmetry of this Hamiltonian under
parity reflection of $J_x \rightarrow -J_x$.  The dynamics of this two
mode system are greatly simplified by this symmetry.

The operator $\hat{J}_{x}$ represents the difference in atom number
between wells 2 and 1 while the total number of atoms is given by
$c_1^\dagger c_{1}+c_2^\dagger c_{2}=N_1+N_2=N$.  As a consequence of
number conservation, the operators of
Eqs.(\ref{jjj1},\ref{jjj2},\ref{jjj3}) have a Casimir invariant given
by
\begin{equation}
\hat{J}^{2}=\frac{\hat{N}}{2}(\frac{\hat{N}}{2}+1).
\label{casimir}
\end{equation}
$\hat{J}_{y}$ and $\hat{J}_{z}$ are of no particular significance for our
present purposes and are not considered further.

\section{The Equations of Motion}
\label{eqofm}

In reference \cite{milburn1997} the mean-field equivalent of the two
mode approximation used here was shown to result from a moment
factorization ansatz in the Heisenberg equations of motion. In effect
this is equivalent to a classical two mode approximation  for the
Gross-Pitaevskii equation.  We start by using the Heisenberg equation
of motion to find the time evolution of the three operators in
Eqs.(\ref{jjj1},\ref{jjj2},\ref{jjj3}).  Using the familiar commutation
relations $[\hat{J}_i,\hat{J}_j]=-i\hat{J}_k$, we obtain three
equations of motion for the operators $\hat{J}_x, \hat{J}_y$ and
$\hat{J}_z$.
\begin{equation}
\frac{d\hat{J}_x}{dt}=-\Omega(t)\hat{J}_y ,
\label{heisenberg1}
\end{equation}
\begin{equation}
\frac{d\hat{J}_y}{dt}=\Omega(t)\hat{J}_x-2\kappa (\hat{J}_z
\hat{J}_x+\hat{J}_x\hat{J}_z) ,
\label{heisenberg2}
\end{equation}
\begin{equation}
\frac{d\hat{J}_z}{dt}=2\kappa(\hat{J}_y\hat{J}_x+\hat{J}_x\hat{J}_y) .
\label{heisenberg3}
\end{equation}
We transform these Heisenberg equations of motion into a
semiclassical system of equations by taking the expectation value
of the above equations and factorizing operator products such as $
\langle\hat{J}_x\hat{J}_z  \rangle=\langle \hat{J}_x\rangle
\langle \hat{J}_z\rangle$.  This leaves us with classical numbers
in place of the $\hat{J}$ operators.  We make the identification
$S_i=\frac{\langle \hat{J}_i \rangle}{N}$, i.e. we scale the
$\hat{J}$ equations by the total number of atoms N.  We now write
down the semiclassical system of scaled equations,
\begin{eqnarray}
\dot{S}_{x}&=&-\Omega(t) S_{y}, \label{sx} \\
\dot{S}_{y}&=&\Omega(t) S_{x}-4\kappa NS_{x}S_{z}, \label{sy} \\
\dot{S}_{z}&=&4\kappa NS_{x}S_{y}, \label{sz}
\end{eqnarray}
with the usual constraint to the sphere,
\begin{equation}
S_{x}^{2}+S_{y}^{2}+S_{z}^{2}=1/4=S^{2}.
\label{sss}
\end{equation}
We identify three real variables $S_k$, corresponding to the operators
$\hat{J}_k$,
\begin{equation}
S_{x}=\frac{1}{2}( b_2^*b_2-b_1^*b_1)
\label{sone}
\end{equation}
\begin{equation}
S_{y}=\frac{-i}{2}( b_{1}^{*}b_{2}-b_{1}b_{2}^{*})
\label{stwo}
\end{equation}
\begin{equation}
S_{z}=\frac{1}{2}(b_{1}^{*}b_{2}+b_{1}b_{2}^{*})
\label{sthree}
\end{equation}
where the $b_d$'s are c-numbers, and are the expectation values of
the second quantized annihilation and creation operators $c_d$.
These real variables $S_k$ can be shown to represent the following
physical parameters, $NS_x$ represents the difference in atom
number between wells 2 and 1, $NS_y$ represents the momentum of the
condensate and $NS_z$ represents the population difference between
the global symmetric and antisymmetric modes of the confining
potential~\cite{milburn1997}.

\section{Semiclassical Dynamics}
\label{semiclassicaldynamics}

\begin{figure}
\begin{center}

\includegraphics[width=85mm,keepaspectratio]{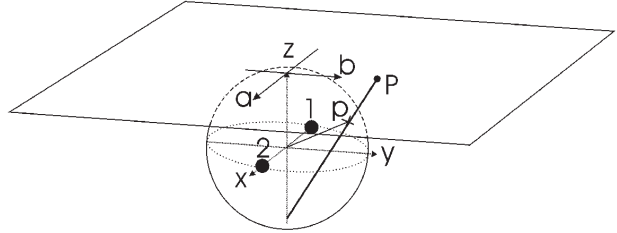}

\end{center}

\caption{Diagram illustrating construction of the stereographic
projection from a sphere.  Points 1 and 2 represent the positions of
the two wells that constitute the double well potential we are
modeling.  p is a point on the sphere projected to the point P on the
plane.  a and b are the coordinates in the plane, as defined in the
text.}
\label{fig_one}
\end{figure}

While we can imagine a vector tracing out some trajectory on the
surface of the sphere, see Fig.~\ref{fig_one}., it is convenient to make a
stereographic projection of the dynamics onto a two dimensional plane.


We use a stereographic projection of a sphere onto the plane $z=1/2$
via the transformation,
\begin{equation}
\nu =\frac{S_x+iS_y}{\scriptstyle 1/2 \displaystyle + S_z}.
\label{varplus}
\end{equation}
Using the definitions $a=Re(\nu)$ and $b=Im(\nu)$ we write,
\begin{equation}
a=\frac{S_x}{\scriptstyle 1/2 \displaystyle + S_z},
 \label{nua}
\end{equation}
\begin{equation}
b=\frac{S_y}{\scriptstyle 1/2 \displaystyle + S_z}.
\label{nub}
\end{equation}

Setting $\lambda=\kappa N/\Omega_0$, we rewrite the
system of Eqs.(\ref{sx},\ref{sy},\ref{sz}) as,
\begin{eqnarray}
\dot{a}&=&-(1+\epsilon cos\frac{\omega_D}{\Omega_0}t) b - 4\lambda \frac{a^2 b}
{1+a^2 + b^2} \; , \nonumber \\
\dot{b}&=&(1+\epsilon cosfrac{\omega_D}{\Omega_0}t) a - 2\lambda a
\frac{1-a^2 + b^2}
{1+a^2 + b^2} \; . \label{realimagnu}
\end{eqnarray}
Where we have scaled the time $t$ by $\Omega_0$ and we set
$\omega_D=1.37\Omega_0$.
The equator of the sphere is mapped to the unit circle in (a,b)
phase space due to our choice of radius ($r=\frac{1}{2}$) for the
sphere and the position of the wells are mapped to $(\pm 1,0)$.

For $\epsilon=0$, i.e. $\Omega(t)=\Omega_0$, the system
(\ref{realimagnu}) is integrable.  If $\lambda <1/2$, the only feature is
a center at the origin.  At $\lambda=1/2$, the origin
undergoes a pitchfork bifurcation creating two stable centers, which
for $\lambda>1/2$ have position
\begin{equation}
a=\sqrt{\frac{2\lambda-1}{2\lambda+1}}, \; \; \   b=0,
\label{bifparam}
\end{equation}
and are surrounded by a `figure 8' type separatrix.
The integral curves for this system are,
\begin{equation}
c(a,b)=\frac{(1+a^2+b^2)^2}{(1+a^2+b^2)+2\lambda a^2}.
\label{intcurves}
\end{equation}
On the separatrix $c=1$, for $c<1$
solutions to Eq.(\ref{intcurves}) lie inside the separatrix and for $c>1$
they lie
outside.

\begin{figure}
\begin{center}

\includegraphics[width=85mm,keepaspectratio]{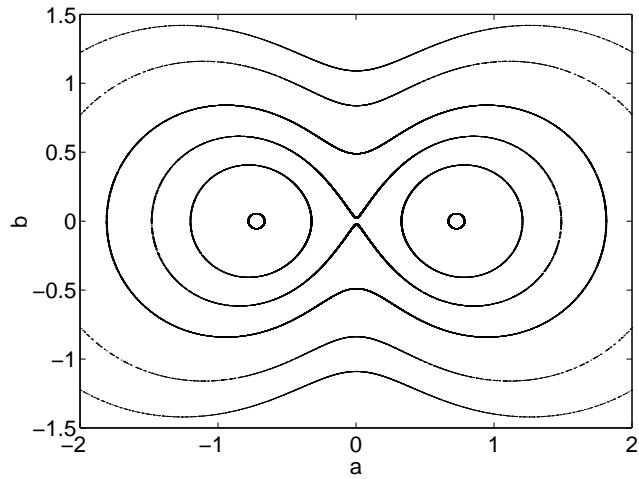}

\end{center}

\caption{Phase space plot of the semiclassical dynamics showing period-1
fixed points and separatrix.  $\epsilon=0.0$, $\lambda=1.6$.}
\label{fig_two}
\end{figure}

Fig.~\ref{fig_two}. shows the phase plane for $\epsilon=0$,
illustrating the integral curves.  If we consider starting in one well,
say $(a=+1,b=0)$, then for $\lambda < 1$, complete population
oscillations between the wells occur~\cite{milburn1997}. At $\lambda
=1$ we find the separatrix intersects the centers of the wells and
complete oscillations no longer occur.  As $\lambda $ increases beyond
$1$ (larger $N$, i.e. more atoms) the solutions of
Eqs.(\ref{realimagnu}) become localized near the fixed points of the
phase plane.  This is known as self-trapping.  Physically, this can be
interpreted as an increase in the self energy with particle number
which modifies the effective single particle potential of the system
and reduces the single particle tunneling~\cite{milburn1997}.

A very similar model with harmonic driving is considered by Elyutin and
Rogovenko~\cite{elyutin}.  In that paper they find fixed points and a
separatrix for their model analagous to the fixed points and separatrix
in the model treated here.  They consider the break down of the regular
phase space due to the overlap of resonances, and show tunnelling
between the classical fixed points of the semiclassical phase space. 
We go a step further here in that direction firstly to analyze the
dynamics of fixed points of the Poincar\'e map (resonances) and initial
conditions localized on those semiclassical resonances and secondly to
analyze a full quantum description.

\subsection*{Resonances}
\label{resonances}

\begin{figure}
\begin{center}

\includegraphics[width=85mm,keepaspectratio]{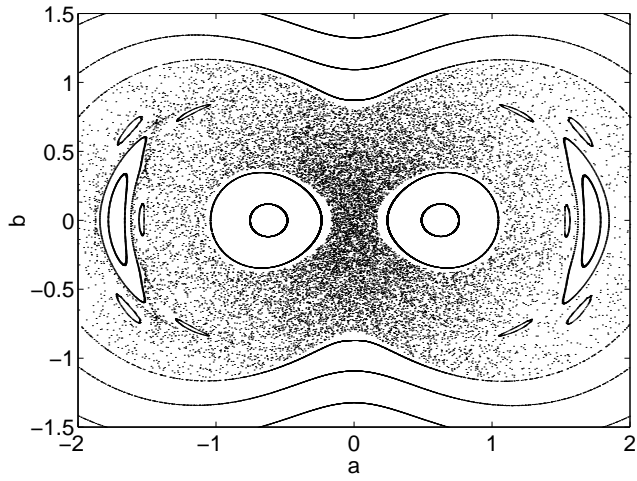}

\end{center}

\caption{Phase space plot of semiclassical driven dynamics showing
period-1 resonances at $a=\pm 1.72,0.0$.  $\epsilon=0.3$, $\lambda=1.6$,
$\omega_D=1.37$.}
\label{fig_three}
\end{figure}

For $\epsilon>0$ the modulated tunneling rate $\Omega(t)$ gives rise to
resonances.  Outside the separatrix, a resonance will appear when there
is a rational ratio between the driving period
and the period of motion of the system on a given integral curve.  We
know on a particular integral curve $c(a,b)$, the system period is
given by,
\begin{equation}
T_{sys}(c,\lambda)=\frac{4K(k)}{\sqrt{w_+-w_-}}, \label{periodof}
\end{equation}
where $K(k)$ is a complete elliptic integral of the
first kind, $k^2=\frac{-2\lambda-w_-}{w_+-w_-}$ and $w_+, w_-$ are
given by
\begin{equation}
w_\pm= \frac{-(2\lambda-1)\pm\sqrt{(2\lambda-1)^2+8\lambda (c-1)/c}}{2}.
\label{wroots}
\end{equation}
In particular, the integral curve given by
$T_{sys}(c,\lambda)=\frac{2\pi \Omega_0}{\omega_D}$ is resonant, breaking up
under perturbation to give two stable period-1 resonances which for
$t=0$ lie on the momentum axis.  These can be seen in the Poincar\'{e}
section, $t=0~ mod(\frac{2\pi \Omega_0}{\omega_D})$, which here is simply a
stroboscopic map taken at
$t_n=n\frac{2\pi \Omega_0}{\omega_D}$. In such a stroboscopic map, see
Fig.~\ref{fig_three}., the
resonances appear as fixed points.  They lie outside the
separatrix, which has been replaced by a `chaotic sea', and are distinct
from the two period-1 fixed points nearer the origin.
For $ \frac{ \omega_D}{ \Omega_0} >1$, here it was taken as 1.37,
these period-1 resonances exist for
$ \lambda > ( \frac{\omega_D}{\Omega_0})^2-1)/2$.


In order to make a comparison with the quantum dynamics we need to
consider an initial classical distribution rather than a single point,
and compute average values of the dynamical variables as the
distribution evolves in time.
The classical phase space
density (distribution) we require to
simulate the quantum state must correspond to a system of points
localized on the spherical classical phase space and should describe the
optimal simultaneous determination of the phase space variables, that is
the components of angular momentum.  These states are known
as the SU(2) coherent states~\cite{Appleby2000} and are defined in
Eq.(\ref{atomiccoherentstate}).

\begin{figure}
\begin{center}

\includegraphics[width=85mm,keepaspectratio]{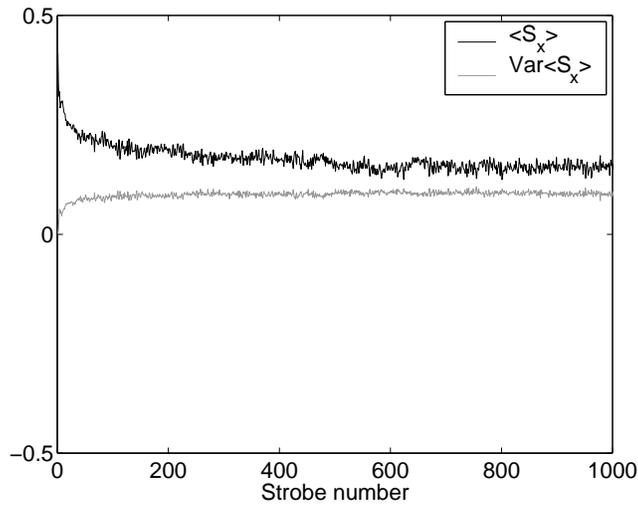}

\end{center}

\caption{Plot of the weighted mean of semiclassical population
difference $<S_x>$ and variance $Var<S_x>$, with initial distribution
centered on the resonance $a=1.72, b=0.0$.  $\epsilon=0.3$,
$\lambda=1.6$, $\omega_D=1.37$.}
\label{fig_four}
\end{figure}

For a quantum state
$|\psi\rangle$, the distribution that describes the output for optimal
simultaneous measurement of angular momentum is
\begin{equation}
P(\alpha)=|\langle\alpha|\psi\rangle|^2
\end{equation}
where $\alpha=e^{i\phi}\tan(\frac{\theta}{2})$. If the system begins in an
initial SU(2) coherent state $|\alpha_0\rangle$,
the corresponding distribution is~\cite{bsanders}
\begin{equation}
P(\alpha|\alpha_0)=\left
[\frac{(1+\alpha_0\alpha^*)(1+\alpha_0^*\alpha)}{(1+|\alpha_0|^2)(1+|\alpha|^2)}
\right ]^{2j},
\end{equation}
where $j$ is the total angular momentum quantum number. This is the
classical distribution that we sample to compute the ensemble dynamics.
The distribution is normalized with respect to the spherical
integration measure, which in the stereographic coordinates is
$d^2\alpha(1+|\alpha|^2)^{-1}$. If we start an initial distribution of
points in a stable resonance island, the population remains confined by the
invariant tori bounding the island, see Fig.~\ref{fig_four}.  In the
next section we shall see how the equivalent quantum simulations
produce a very different result.


\section{Quantum Dynamics}
\label{qm}

It is difficult to solve nonlinear Heisenberg equations of motion so we
work in the Schr\"odinger picture using an orthonormal basis defined
by the eigenstates of $J_z$, as in~\cite{milburn1997}.  Using the familiar
angular momentum notation we write our state $|\psi\rangle$ as,
\begin{equation}
|\psi\rangle=\sum_{n=-j}^j c_n(t)|j,n\rangle
\label{jzbasis}
\end{equation}
and substitute into the Schr\"{o}dinger equation to get
\begin{equation}
i\hbar\dot{c}_n(t)=\sum_{m=-j}^j\langle j,n|H|j,m\rangle c_m(t).
\label{quantumequation}
\end{equation}
This is a set of linear coupled equations ($2j+1$ of them) and we use a
Runge-Kutta numerical routine to solve them for the coefficients
$c_n(t)$. Given these coefficients we can obtain the various expectation
values $\langle \hat{J}_i\rangle$ and thus the solution of our model.  We once
again treat the driven and undriven systems separately.
For $\epsilon=0$ we obtain regular collapse and revival sequences
as shown previously in~\cite{milburn1997}.

For $\epsilon>0$ the system is described by a time periodic Hamiltonian. In
that
case the appropriate description is in terms of the Floquet operator $\hat{F}$,
of the system which maps the state from one time to a
time exactly one modulation period later~\cite{haake},
\begin{equation}
|\psi^{(n+1)}\rangle = \hat{F} |\psi^{(n)}\rangle \;
n=0,1,2,\dots\ .
\label{floquet}
\end{equation}
The Floquet map is the quantum equivalent of the classical Poincar\'{e}
section defined in section~\ref{semiclassicaldynamics}.  We obtain the
Floquet Operator in the basis which diagonalizes $\hat{J}_z$ as
follows.  Solve the Schr\"{o}dinger equation over one modulation period
of time $T$ to obtain $|\psi_m(T)\rangle$ for the set of initial
conditions $\{|j,m\rangle\, m=-j,-j+1,\ldots,j-1,j\}$ . The solution
vectors, $|\psi_m(T)\rangle$, form the columns of the Floquet Operator
matrix $\hat{F}$.

\begin{figure}
\begin{center}

\includegraphics[width=85mm,keepaspectratio]{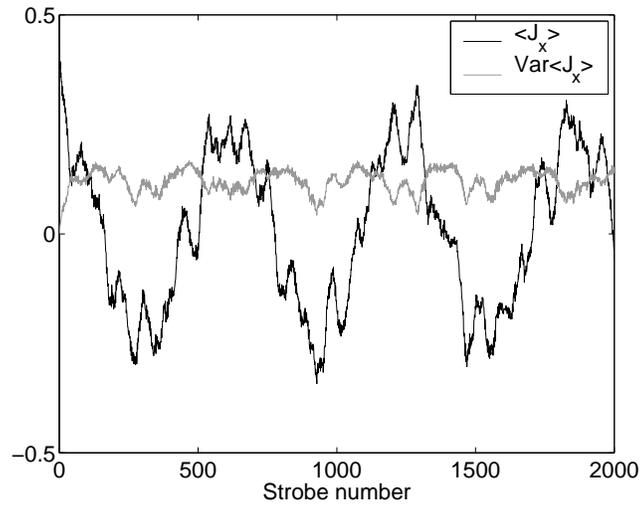}

\end{center}

\caption{Plot of quantum model population difference $<J_x>$ and
variance $Var<J_x>$, with initial atomic coherent state centered on the
resonance $a=1.72, b=0.0$. $\epsilon=0.3$, $\lambda=1.6$, $N=100$,
$\omega_D=1.37$.}
\label{fig_five}
\end{figure}

In  Fig.~\ref{fig_five}., we plot the mean value of $\hat{J}_x$ for an
initial state localized on the same period-one resonance as in
Fig.~\ref{fig_four}.
We use the SU(2) coherent states as the initial localized states.
These are defined on the sphere by~\cite{milburn1997},
\begin{equation}
\ | \alpha \rangle=\sum_{m=-j}^{j}\left( \begin{array}{c}{2j} \\ {m+j}
\end{array} \right)^{1/2} \frac{ \alpha^{m+j}}{(1+ | \alpha |^{2})^{j}}
\ |j,m\rangle .
\label{atomiccoherentstate}
\end{equation}
where $\alpha=e^{i\phi}tan(\frac{\theta}{2})$, and $\theta$ and $\phi$
are the spherical polar coordinates.  For the parameters $\epsilon=0.3$
$\lambda=1.6$ it is found that after $310$ modulation periods,  the
system state appears to be localized on the other resonance, see
Fig.~\ref{fig_five}.  This behavior is known as {\em dynamical
tunneling}~\cite{Davis1981}. Dynamical tunneling is a different process
from the well-known case of barrier tunneling. There is no energetic or
potential barrier. None the less in dynamical tunneling the system
exhibits classically forbidden motion, motion which cannot be accessed
by a classical particle moving on any trajectory.
Dynamical tunneling in a single atom system has recently been observed
by two groups~\cite{Raizen,Hensinger}.

We can understand the tunneling in the following way.  The Hamiltonian
is invariant under rotations about the z-axis, i.e. operations of
$e^{i\pi\hat{J}_z}$ leave the Hamiltonian unchanged. Thus the Floquet
states, by the nature of their construction must fall into two sets,
corresponding to the eigenvalues of $e^{i\pi\hat{J}_z}$.  These two
sets contain vectors of odd and even parity.

Let  $|\psi_-\rangle$ be a state localized on
one of the period-one resonances with coordinates $(-1.72,0)$, while
$|\psi_+\rangle$, is localized on the other period one resonance,
$(+1.72,0)$.
These states are related by a rotation of $|\psi_-\rangle$ by $\pm \pi$
about the $z$ axis,
\begin{equation}
e^{i\pi J_z}|\psi_+\rangle=|\psi_-\rangle.
\label{rotation}
\end{equation}
As the Hamiltonian is invariant under rotations of $\pi$ around the z
axis, the eigenstates of the Floquet operator fall into two parity
classes defined by this rotation. If the initial state is well
localized on one of the period-one resonances, it will tend to have
maximal support in a two dimensional subspace spanned by two particular
Floquet eigenstates of opposite parity.   We label these two simultaneous
eigenstates of $\hat{F}$ and parity as $|\phi_\pm\rangle$, where
$e^{i\phi_\pm}$ are the corresponding eigenvalues of the Floquet
operator. We then write the `negative' and `positive' localized states,
$|\psi_{\pm}\rangle$ as a superposition of these parity eigenstates
states, i.e.
\begin{equation}
|\psi_+\rangle=k_1|\phi_+\rangle+k_2|\phi_-\rangle,
\label{evenparity}
\end{equation}
and,
\begin{equation}
|\psi_-\rangle=k_1|\phi_+\rangle-k_2|\phi_-\rangle,
\label{oddparity}
\end{equation}
where $k_1,k_2$ are very nearly equal and
together almost exhaust the normalization condition, $k_1^2+k_2^2\approx 1$.

For complete tunneling we must have,
$|\psi_+\rangle \rightarrow  |\psi_-\rangle$.
How many iterations of Eq.(\ref{floquet}) satisfy this tunneling condition?
Suppose we start in the state $|\psi_{+}\rangle$.  After $n$ iterations of
the Floquet map we reach the state
\begin{equation}
|\psi^{n}\rangle=\hat{F}^n
|\psi_{+}\rangle=\left [ \frac{1}{\sqrt{2}}(e^{in\phi_+
}|\phi_+\rangle+e^{in \phi_-}|\phi_-\rangle ) \right].
\label{floquet2}
\end{equation}
When the relative phase of the two states in this superposition is
shifted by $\pi$, the state becomes almost equal to a state localized
in the other period-one resonance $|\psi_{-}\rangle$. This determines
the tunneling time as
\begin{equation}
n_t=\frac{\pi}{\phi_--\phi_+}.
\end{equation}
To find the particular Floquet eigenstates on which an initially localized
state has maximum support we simply expand the initial state
$|\psi_{+}\rangle$ in the basis of Floquet eigenstates. This also gives
the corresponding eigenphases and thus the tunneling time. By this
method we find the tunneling time should be $n_t= 310$. This compares
favorably with the tunneling time of about $n= 313$ evident in
Fig.~\ref{fig_five}.

We expect to see the variance of $\hat{J}_x$ at a minimum whenever the
$\hat{J}_x$ component is most localized around each of the resonances,
which is when the state should be close to a minimum uncertainty state.
At the half way point, a coherent superposition of two states of equal
and opposite mean values of $\hat{J}_x$ should occur, and the variance
should be a maximum.  This behavior is evident in Fig.~\ref{fig_five}.

\section{Conclusion}
\label{conclusion}

In this paper we have used a two mode approximation to describe the
quantum dynamics of a BEC in a double well potential with periodic
modulation of the potential barrier separating the two wells. In
mean-field theory (Gross-Pitaevskii limit) we find, for appropriate
parameter values, a mixed phase space of regular stable motion,
associated with fixed points, coexisting with chaotic dynamics. We have
shown the existence of dynamical quantum tunneling between period-one
fixed points (resonances).  This dynamical tunneling is distinct from
the single particle tunneling present in the undriven system and may be
interpreted as a new signature of the quantum dynamics of a nonlinear
quantum field.

Double well potentials similar to those modeled here have been
experimentally realized, see for
example~\cite{Andrews1997,Raizen,daviesetal}. While current experiments
are unlikely to be well described by the two mode model given here, we
expect dynamical tunneling to be sufficiently robust that an
experimental realization of the dynamical quantum tunneling predicted
here can be observed. The recent observation by the NIST
group~\cite{Hensinger} of dynamical tunneling of a single particle
system in a modulated periodic potential in fact used a BEC, although
not in a regime where the mean-field is significant.  It would not be
difficult to modify the experiment so that the nonlinear effect of the
BEC would be significant.

\section*{Acknowledgements}

G.L.S. wishes to thank Kae Nemoto and Bill Munro for very helpful discussions.





\end{document}